\documentstyle[11pt,newpasp,twoside,epsf]{article}
\def\edcomment#1{\iffalse\marginpar{\raggedright\sl#1\/}\else\relax\fi}
\reversemarginpar

\begin{document}
\title{First results from multi-wavelength observations during the 2001 outburst of WZ\,Sge}

\author{Erik Kuulkers} 

\affil{SRON, National Institute for Space Research, Sorbonnelaan 2, 3584 CA Utrecht \&\
Astronomical Institute, Utrecht University, P.O.\ Box 80000, 3508 TA Utrecht, 
The Netherlands}

\author{Christian Knigge, Danny Steeghs} 

\affil{Department of Physics \&\ Astronomy, University of Southampton, 
Highfield, Southampton, SO17 1BJ, UK}

\author{Peter J.\ Wheatley}

\affil{Department of Physics \&\ Astronomy, University of Leicester, University Road, Leicester, LE1 7RH, UK}

\author{Knox S.\ Long}

\affil{Space Telescope Science Institute, 3700 San Martin Drive, Baltimore, MD 21218, USA}

\author{On behalf of a larger collaboration}

\begin{abstract}
WZ\,Sge has shown superoutbursts in 1913, 1946 and 1978. On 2001 July 23 a
new outburst was announced, about 10 years `too early'. Target of opportunity 
satellite observations with Chandra, FUSE, HST and RXTE were performed throughout the
outburst. From the ground WZ\,Sge was monitored by numerous professional and
amateur astronomers, in the optical, IR and radio. 
We give an account of the first exciting results from these multi-wavelength observations.
\end{abstract}

\section{Introduction}

The cataclysmic variable (CV) 
WZ\,Sge was classified as a recurrent nova after it had displayed the outbursts
in 1913 and 1946 (Mayall 1947). However, subsequent optical observations showed the source
to be a dwarf nova (e.g.\ Warner 1976). Its class was finally established
during the outburst in 1978, when the source displayed superhumps at a period 
which was about 0.8\%\ longer than the orbital period, thus showing it to be
an SU\,UMa star (e.g.\ Patterson et al.\ 1981). 
WZ\,Sge, in fact, belongs to the `extreme' SU\,UMa
subclass named after itself, i.e.\ the WZ\,Sge stars. They exhibit very
infrequent and large amplitude [super]outbursts (e.g.\ Bailey 1979).

Mass estimates of the white dwarf range from 0.25--1.2\,M$_{\odot}$ (e.g.\ Patterson 1998).
During the 2001 outburst, Steeghs et al.\ (2001b) detected the secondary (see Sect.~3.2)
and found that the donor has a relatively low mass of $<$0.1\,M$_{\odot}$ 
(see also Patterson 2001). The derived mass function implies that the white dwarf has a mass 
$>$0.77\,M$_{\odot}$. 
Binary evolution models predict that all CVs should eventually evolve 
through a minimum orbital period around 70\,min, when the mass donor becomes 
degenerate. With an orbital period of 81.6\,min (Krzemi\'nski 1962; see Skidmore et al.\ 
2000, for a latest update), WZ\,Sge is close to that minimum period.
The inclination at which we view the source is 75$^{\circ}$, just high enough for
the donor to obscure the accretion disk but not the white dwarf
(Krzemi\'nski 1962; Smak 1993). 
The white dwarf rotates rapidly (Cheng et al.\ 1997) at a period near 28\,s 
(Patterson et al.\ 1998; Lasota, Kuulkers \&\ Charles 1999, and references therein; 
but see Skidmore et al.\ 1999).

At the distance of 45\,pc (astrometric parallax, J.~Thorstensen 2001, priv.\ comm.) 
it is one of the closest CVs.
When in outburst it is therefore bright (up to 7.5--8\,mag), providing an excellent
target to investigate the accretion of matter from the donor, through the disk, onto 
the white dwarf, not only for professionals but also for amateur observers.

\section{The 2001 outburst}

On Monday 2001 July 23, the astronomical community was  alerted by the Variable Star Network
(VSNET)\footnote{http://www.kusastro.kyoto-u.ac.jp/vsnet/Mail/alert6000/msg00093.html}
that WZ\,Sge had brightened considerably by $\sim$5.5\,mag 
from its quiescent magnitude of $\sim$15.5 (Ishioka et al.\ 2001a; 
Mattei 2001).
Based on the times of three previous outbursts a new outburst had not been expected until 2011.
The 2001 outburst therefore seemed to be 10 years `too early'.
However, we note that although the superoutburst recurrence times of SU\,UMa stars 
are quite stable, they do change from time to time (Vogt 1980). 

\begin{figure}
\plotfiddle{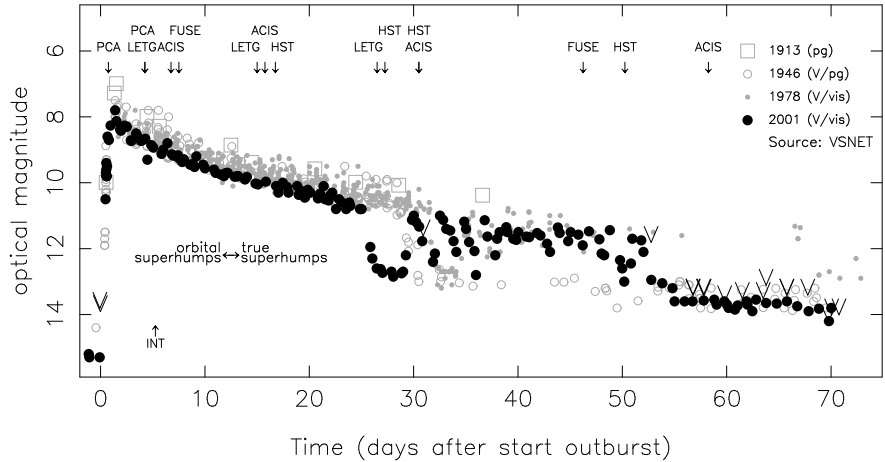}{6cm}{-90}{50}{50}{-200}{220}
\caption{Optical observations of the 2001 outburst as collected by VSNET.
Data after maximum are 0.25\,day averages. 
Upper limits are indicated with $\bigvee$. 
At the top we indicated the times of observations performed
with various instruments/satellites. The PCA is onboard RXTE, whereas the LETG and
ACIS are onboard Chandra. 
At the bottom we have indicated the time of the 
INT observations showing spiral arms (see Fig.~2). 
For comparison we show in light gray the outbursts in 1913, 1946 and 1978 (from Kuulkers 1999).}
\end{figure}

In Fig.~1 we show the visual and other optical observations of the 2001 outburst 
as reported to VSNET\footnote{For an excellent optical coverage and many examples of light curves
we refer to the VSNET webpage dedicated to WZ\,Sge:
{\tt http://www.kusastro.kyoto-u.ac.jp/vsnet/DNe/wzsge01.html}.} (see also Ishioka et al.\ 2001b).
Within half a day WZ\,Sge brightened from its quiescence level
to about 10.5\,mag. It peaked at 8\,mag, giving a large outburst amplitude 
of about 7.5\,mag. The main outburst lasted for about 25~days, after which it displayed 
a dip of 2\,mag lasting for about 5~days. It then showed irregular 
rebrightening/outburst behaviour for another month. It then declined to 13.5\,mag; since then
no new rebrightening has been observed, and the superoutburst seems to have ended.

Also in Fig.~1 we have overplotted in gray the previous observed outbursts (visual, photometric and
photographic measurements; from Kuulkers 1999). Clearly, the 2001 outburst resembles the 1978
outburst, except for the fact that the durations of the main outburst and the outburst stage after 
the dip were shorter (about 5~days and 10~days, respectively) for the 2001 outburst. Note that
the brightness after the 2001 outburst is still about 2\,mag above quiescence, at the same level as 
after the 1946 outburst. This is due to the hot white dwarf, see Sect.~3.4.

Both `orbital superhumps' (Kato et al. 2001a), commonly
observed in the beginning of the superoutbursts of 
WZ\,Sge-stars, and `true' (Kato et al.\ 2001b) superhumps were present. 
The true superhumps appeared about two weeks into the outburst, whereas
at the same time the orbital superhumps died away. Note that the change from orbital to true
superhumps is not abrupt. They overlap and coexist in the light curves for a few days, 
suggesting a different nature.

The superoutburst of WZ\,Sge was extensively monitored from the ground and from 
space\footnote{See {\tt http://www.astro.soton.ac.uk/$\sim$ds/wzsge.html}.}. 
In the next sections we give an account of the first results from 
(mainly spectroscopic) observations at optical, UV and X-ray 
wavelengths in which we have been involved,
as well as a few radio observations.

\section{Multi-wavelength observations during the 2001 outburst}

\subsection{Radio observations: no detection}

Various radio observations 
(VLA: 3.6\,cm, 6\,cm; V.~Dhawan, M.~Rupen 2001, priv.\ comm.; 
Ryle: 2\,cm; G.~Pooley 2001, priv.\ comm.)
were performed between July 24 and Aug 9, but yielded no detections.
Upper limits on e.g.\ Aug 4 were 0.12 and 0.18\,mJy per beam at 3.7 and 6\,cm,
respectively (M.~Rupen 2001, priv.\ comm.).

\subsection{Optical spectroscopy: spiral waves and secondary star}

One of the first (July 23.74) optical spectra showed
H$\alpha$ and H$\beta$ lines in absorption on top of a blue continuum
(Ishioka et al.\ 2001a). About a day later the spectra had changed dramatically,
now showing strong double-peaked emission lines of e.g.\ He\,II (4686\AA\/)
and C\,III and N\,III. H$\beta$ showed both emission and absorption features
similar to P\,Cyg profiles, whereas higher series were seen as broad absorption
features. The emission profiles showed variations on the 
orbital period (Baba, Sadakane \&\ Norimoto 2001). 

\begin{figure}
\plotfiddle{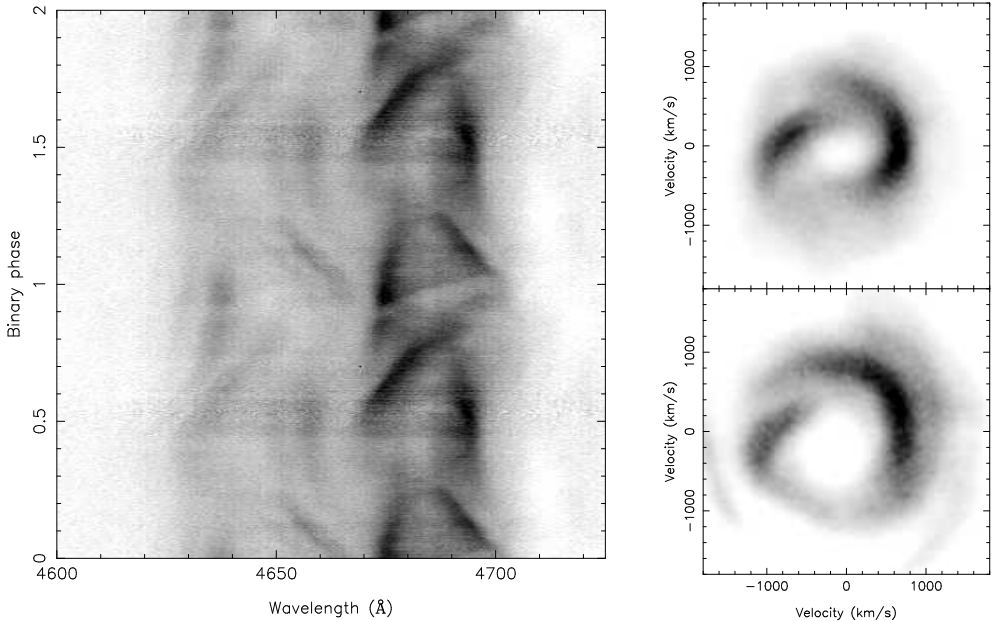}{8.5cm}{0}{87}{87}{-300}{-260}
\caption{{\it Left}: The observed emission line profiles of He\,II (4686\,\AA\/) and the Bowen
complex on Aug 28, 5 days into   the outburst. Two  orbital
cycles are plotted.
{\it Right}: Corresponding Doppler tomograms of   He\,II (top) and C\,III  reveal
prominent two-armed spirals dominating the emission from the accretion disk.}
\end{figure}

Service override observations on the 2.5\,m INT at La Palma (PI: Steeghs) were performed
between July 28.02--28.24 (see Fig.~1), yielding high resolution phase resolved spectra
(Steeghs et al.\ 2001a).
The same emission lines as seen by Baba et al.\ (2001) were present.
In Fig.~2 we  display the phase  dependent emission profiles of He\,II
(4686\,\AA\/) and the nearby  Bowen fluorescence complex.  The Bowen blend
consists of several C\,III and N\,III transitions, but is dominated by C\,III
and closely  reflects the line kinematics  of  He\,II. The corresponding
Doppler tomograms (Marsh  \&\ Horne 1988) 
are  also plotted in Fig.~2
and reveal   prominent  spiral arms  in   the  accretion disk   of  WZ\,Sge. 
These two armed structures are  very similar to those observed in
longer period  dwarf novae  during  outburst (Steeghs  2001, Boffin \&
Steeghs, these proceedings). 
On the other hand, the Balmer and He\,I lines are dominated by deep, 
phase dependent absorption components.
We conjecture here that the observed spiral arms are
related to the orbital superhumps. 
A more detailed comparison between the observed disk structure
with the photometric behaviour may shed light on this.

Spectroscopy obtained during   later phases of the   outburst revealed
growing narrow   emission components in  the  Balmer lines originating
from the   irradiated secondary star (Steeghs    et al.\  2001b).  Such
emission features from  the secondary  are  commonly observed in   CVs
during outburst, as  well as persistent X-ray binaries  such as Sco\,X-1
(Steeghs  \&  Casares  2001).  For  WZ\,Sge   this 
provided the  first
opportunity to  directly reveal  its low-mass  secondary star.  Apart
from the appearance  of  secondary star  features,  the accretion disk
emission also makes a significant transition  from strong spirals arms
to a disk flow  dominated by an  extended bright  spot around  20 days
into  the outburst. 
The large  amount of optical spectroscopy obtained
during the 2001 outburst of  WZ\,Sge should  shed important insights into 
the  evolution of the accretion  disk.

\subsection{X-ray observations: RXTE and Chandra}

\begin{figure}
\plotfiddle{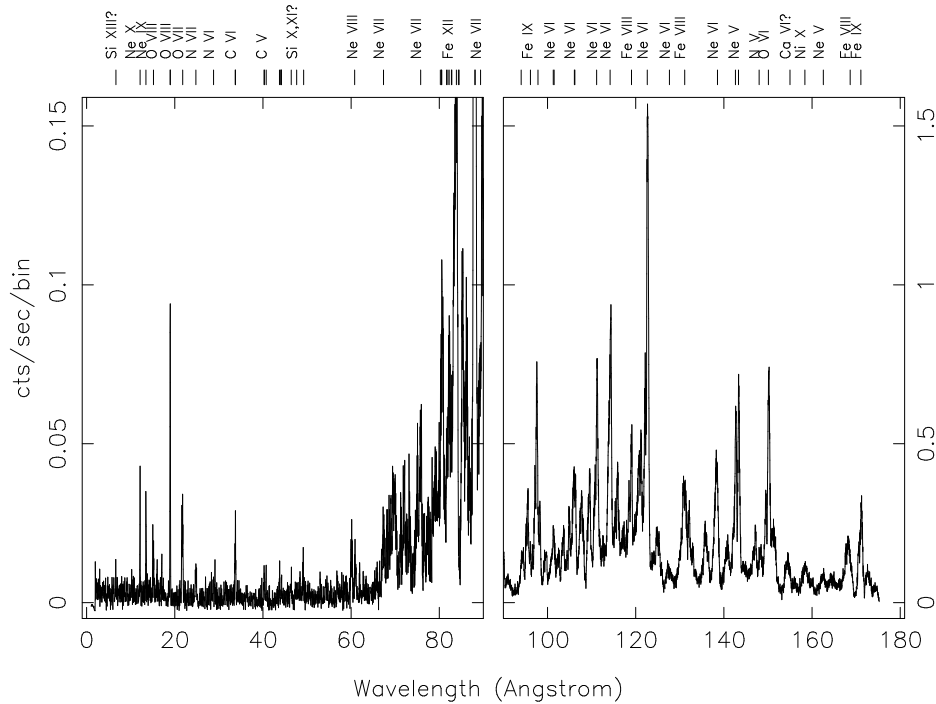}{8.7cm}{-90}{50}{50}{-190}{290}
\caption{Average Chandra/LETGS spectrum obtained on July 27, 3.5 days into the outburst. Note the difference
in scale between the short and long-wavelength part of the spectrum. Indicated are many of the 
line identifications.}
\end{figure}

A 10-min snapshot was performed with RXTE on the same evening of the discovery (UT July 23, 20:40),
followed by a 30-min observation on July 27 06:40 (UT). 
No detections were made with 95\%\ 
confidence upper limits of 5.3$\times$10$^{-12}$\,erg\,cm$^{-2}$\,s$^{-1}$ and
3.7$\times$10$^{-12}$\,erg\,cm$^{-2}$\,s$^{-1}$ (2--6\,keV), respectively, assuming
a typical 6\,keV MEKAL spectrum. We note that due
to the limited sensitivity of RXTE we can not say whether there is an X-ray delay with respect to
the optical outburst, commonly observed in outbursts of dwarf novae.

Three Chandra/LETGS (PI: Wheatley) and four
Chandra/ACIS-S (PI: Kuulkers) observations were performed throughout the outburst, see Fig.~1. 
Note that the second RXTE observation was simultaneous with the 
first Chandra/LETG observation.
XMM-Newton and BeppoSAX were unable to observe WZ\,Sge due to viewing constraints.

The first Chandra/LETGS observation was carried out three and a half days into the outburst
(Wheatley et al.\ 2001). The mean zeroth-order count rate was 4.5\,c\,s$^{-1}$, showing the
source to be bright, and was modulated
with the orbital period. In Fig.~3 we show the averaged 20\,ksec grating spectrum,
which had a mean count rate of 20\,c\,s$^{-1}$. The superb spectrum shows a `forest' of broad
emission lines longward of $\sim$65\AA\ from e.g.\ O\,V--VII, Ne\,V--VIII, Fe\,VII--IX.
The FWHM of these lines are between 800--1200\,km\,s$^{-1}$ (see e.g.\ left panel of Fig.~6). 
At shorter wavelengths ($<$65\,\AA\/)
one sees weaker lines of highly ionized ions, with O\,VIII (18.9\,\AA\/) standing out.
The mean flux in the 2--180\,\AA\ band was 1.6$\times$10$^{-12}$\,erg\,cm$^{-2}$\,s$^{-1}$.

The long wavelength spectrum resembles the EUVE spectra taken during an outburst of
U\,Gem between orbital phase 0.6--0.8 (Long et al.\ 1996),
as well as during a superoutburst of OY\,Car (Mauche \&\ Raymond 2000).
Both U\,Gem and OY\,Car are high-inclination systems, like WZ\,Sge. The forest of broad lines 
may therefore be formed by scattering of emission from the white dwarf's vicinity into the photo-ionized
accretion disk wind. 
Note that some direct optically-thick
boundary layer emission is still visible in the case of U\,Gem.
During the second LETGS observation on Aug 6 this component had faded 
considerably. The mean zeroth-order count rate had dropped to 0.2\,c\,s$^{-1}$.

The first ACIS-S\footnote{Covering 0.2--10\,keV or 1.2--60\,\AA.} 
observation (2.5\,c\,s$^{-1}$) starting on July 29 16:50 (UT) showed the spectrum to be also dominated
by strong emission lines, although they were not as well-resolved as in the LETGS spectrum, due to
the much lower energy resolution. A strong emission line was seen near 
(2.4\,keV, 5.2\,\AA\/), probably associated with He-like S\,XV, which cannot be accounted for
using an optically-thin thermal plasma model.

\subsection{UV observations: FUSE and HST}

\begin{figure}
\plotfiddle{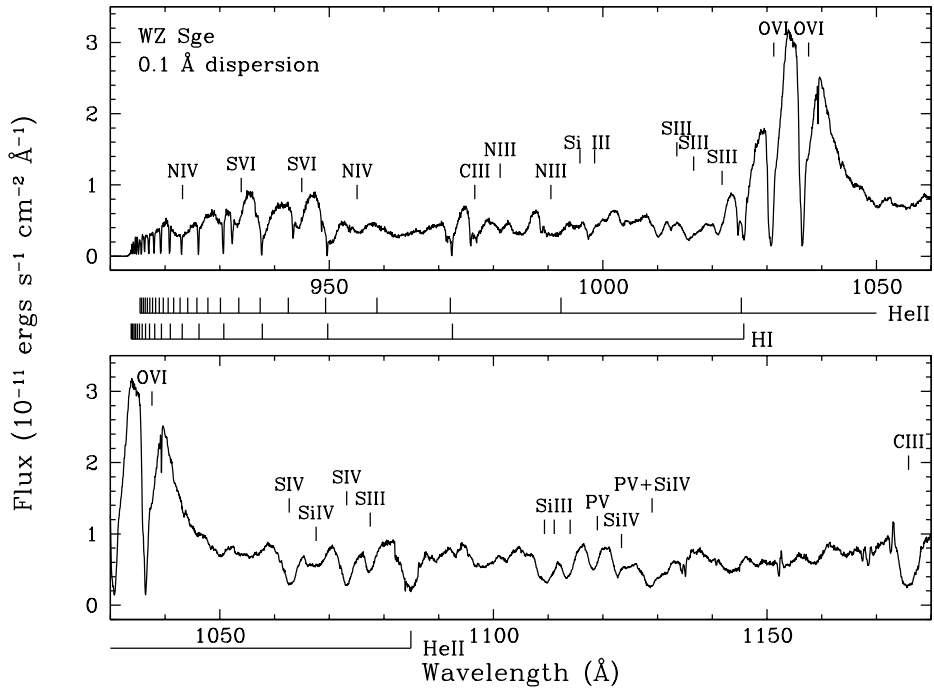}{8.7cm}{-90}{48}{48}{-180}{280}
\caption{Average FUSE spectrum obtained on July 30, 7 days into the outburst. Indicated are the line identifications.}
\end{figure}

The only satellite observations during previous outbursts are those obtained
by IUE in 1978, from the start of the outburst up to the first dip in the
outburst light curve (Fabian et al.\ 1980; Naylor 1989; Smak 1993, and references therein). 

Several FUSE (PI: Long) and HST observations were performed during the 2001 outburst
(see Fig.~1). 
The first HST observations were mainly aimed at 
orbital phase resolved UV spectroscopy (PI: Knigge). The other HST observations (PI: Sion) will continue
way after the outburst, and will (among other things) permit detailed study of 
the cooling white dwarf (see also below).
Note that the third HST observation was simultaneous with the third Chandra/ACIS-S observation.

The mean spectrum of the FUSE observation on July 30 (Fig.~4) 
is dominated by a broad line feature 
due to O\,VI. The EW of
this feature is between 50 and 60\AA, depending on how you treat
the narrow features which punctuate it. Its FWZI is close to 9000\,km\,s$^{-1}$.  
The absorption features which interrupt the emission profile of O\,VI are
blue shifted by about 300\,km\,s$^{-1}$ from the rest wavelengths of the O\,VI
doublet.
The average flux between 1050 and 1150\,\AA\ was 
6$\times$10$^{-12}$\,erg\,cm$^{-2}$\,s$^{-1}$\,\AA$^{-1}$.
The narrow absorption
lines seen in the spectrum are due to interstellar hydrogen.  There is
no evidence of molecular H or of metal absorption lines, which is
expected given WZ\,Sge's proximity.

\begin{figure}
\plotfiddle{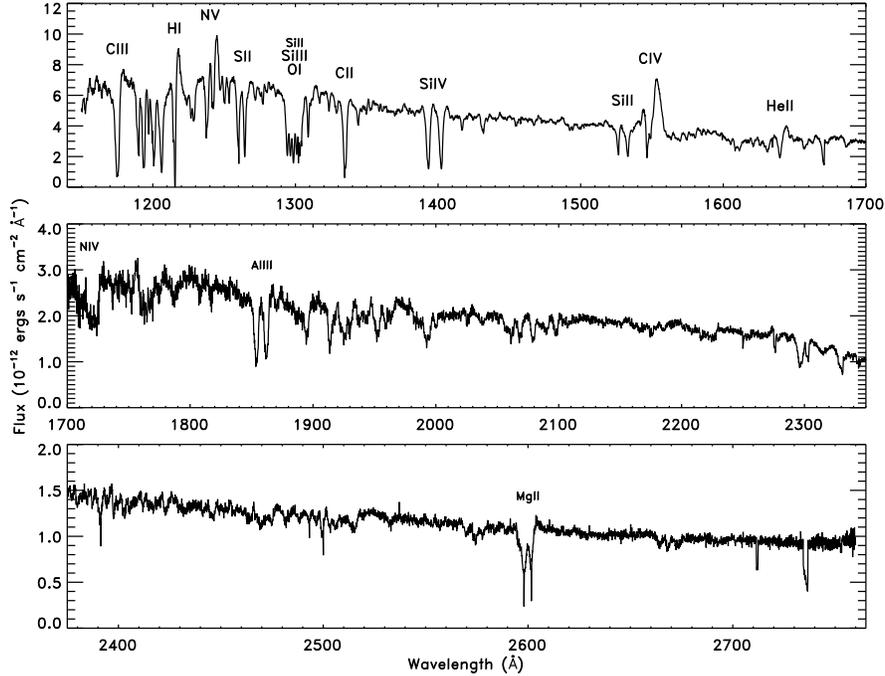}{8.7cm}{90}{55}{55}{230}{-40}
\caption{Average HST/STIS spectrum obtained on Aug 8, 16 days into the outburst. Indicated are some of the
line identifications.}
\end{figure}

This observation preceded the first HST observation taken nine days later
(see below), but in fact the continua of the FUSE and the HST spectrum match up pretty well 
in the overlap region. It is, therefore, reasonable to say that the line spectrum
associated with the 1s--2p lines of Li-like O, N, and C indicates a very
high excitation spectrum.  
The continuum spectrum, on the other hand, is punctuated by a lot of features mostly
associated with lower ionization.  The line widths of the unblended
lines are all of order 500--700\,km\,s$^{-1}$. 
They are somewhat reminiscent of outburst spectra of U Gem obtained with FUSE
(Froning et al.\ 2001; Froning et al., these proceedings).

Fig.~5 shows the mean HST/STIS spectrum obtained on Aug 8;
the average flux between 1150 and 3100\,\AA\ was 
2.3$\times$10$^{-12}$\,erg\,cm$^{-2}$\,s$^{-1}$\,\AA$^{-1}$.
Broadened emission lines 
from N\,V and C\,IV are seen, as well as some He\,II. 
Shortward of 1410\,\AA, a lot of strong broadened absorption lines are present, associated 
with C, S, Si and O. Above 1410\,\AA\ the strongest absorption lines are from 
Al\,III (1850--1860\,\AA\/) and Mg\,II (2790--2810\,\AA\/). The others are most probably 
associated with Fe.
Note that some part of the Mg\,II absorption has an interstellar origin. 
Still, most of it is due to the source, 
since all the other IS lines are weak and WZ\,Sge is rather close.
Note that the spectra obtained by IUE during the 1978 outburst 
(Fabian et al.\ 1980; Naylor 1989) look very similar to the HST spectrum.

In the right panel of Fig.~6 we show the C\,IV profile in more detail. 
The emission is line is clearly broad,
as expected if formed in the wind, and somewhat asymmetric. Superimposed are absorption dips
blueward of the center. 
We note the striking similarity between the C\,IV and N\,VI profiles
(compare both panels of Fig.~6).
The complex C\,IV line profile is somewhat reminiscent of that
seen in the high inclination nova-like variable UX\,UMa (Mason et al.\ 1995).
In that system, the C\,IV line is thought to be
produced in a powerful accretion disk, with the absorption dips arising in
a vertically extended accretion disk chromosphere (Knigge \&\ Drew 1997).
However, it is still far from clear whether the roughly
similar line profiles actually imply physically similar line-forming
regions in WZ\,Sge and UX\,UMa.

As evidenced by IUE spectroscopy of the late phases of the 1978 outburst (Slevinsky et al.\ 1999), 
the slow decline of the optical light curve at the end of each outburst is consistent with cooling of 
the white dwarf. It is interesting to note that the optical brightness at the bottom of the dips in the 
1978 and 2001 outburst light curves fall on the backwards extrapolated white dwarf cooling tracks. 
At such times
not much disk emission would be expected, and the white dwarf would dominate the emission.
Preliminary analysis of the second HST observation during the dip
indicates that the spectrum is indeed dominated by an approximately 28\,000\,K hot white dwarf, 
while the first and third observation resemble that expected from an accretion disk.

\begin{figure}
\plotfiddle{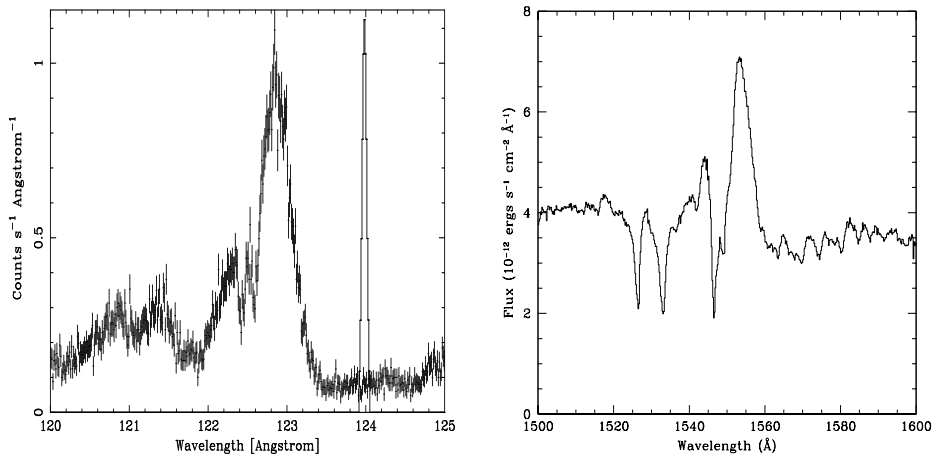}{6cm}{0}{85}{85}{-280}{-285}
\caption{{\it Left}: LETGS spectrum focused on the broad N\,VI emission. 
For comparison we show near 124\,\AA\ 
the energy resolution of the LETGS instrument by a narrow drawn line.
It is not clear yet whether the absorption dips
slightly blueward of N\,VI is due to real absorption or to the presence of two close emission lines.
{\it Right}: HST/STIS spectrum focused on the broad C\,IV emission plus narrow absorption.
The two absorption lines shortward of C\,IV are due to the Si\,II doublet.}
\end{figure}

\section{Conclusions}

It is clear that the 2001 outburst has been looked at in great detail with
various instruments with exquisitely high temporal and high wavelength (or energy) resolution. 
We have shown the first exciting preliminary results of a few of these observations, which already
reveal a rich amount of information. The enormous amount of data accumulated over the outburst
at various wavelength will leave us work for the years to come, and may well reveal the 
nature of superoutbursts in WZ\,Sge stars, as well as SU\,UMa stars in general!

\bigskip
\acknowledgements{The observations shown here represent the work 
of a large collaboration, which includes
T.~Marsh (optical/HST), J.~Casares, E.~Harlaftis, W.~Skidmore (optical), C.~Markwardt, J.~Swank
(RXTE), J.~Drake, J.~Kaastra, C.~Mauche, S.~Starrfield, M.~Wagner 
(Chandra), P.~Szkody, E.~Sion (FUSE), R.~Hynes and R.~Doyle (HST). 
Many thanks to the Isaac Newton Group  staff for obtaining part of the
data through the ING service programme, and V.~Dhawan and M.~Rupen for spending some of their VLA time
on WZ\,Sge.
In addition we are profoundly grateful for the support of many amateur observers who
sacrifice their nights for observing stars like WZ\,Sge, and the
collaboration of organizations such as VSNET and the AAVSO, who enabled
us to plan the observations discussed here.}

\end{document}